\documentclass[a4paper]{jpconf}
\usepackage{graphicx}

\usepackage{amssymb}
\usepackage{amsmath}

\newcommand{\be}{\begin{equation}}
\newcommand{\ee}{\end{equation}}
\newcommand{\beq}{\begin{eqnarray}}
\newcommand{\eeq}{\end{eqnarray}}

\def\lsim{\hbox{ \raise.35ex\rlap{$<$}\lower.6ex\hbox{$\sim$}\ }}
\def\gsim{\hbox{ \raise.35ex\rlap{$>$}\lower.6ex\hbox{$\sim$}\ }}

\newcommand{\bsa}{\begin{subeqnarray}}
\newcommand{\esa}{\end{subeqnarray}}
\newcommand{\bea}{\begin{eqnarray}}
\newcommand{\eea}{\end{eqnarray}}
\newcommand{\ba}{\begin{array}}
\newcommand{\ea}{\end{array}}
\newcommand{\bit}{\begin{itemize}}
\newcommand{\eit}{\end{itemize}}

\begin{document}

\begin{flushleft}
KCL-PH-TH/2013-4 \\
\end{flushleft}

\title{Noncommutative Spectral Geometry: A Short Review
}

\author{Mairi Sakellariadou} \address{Department of Physics, King's
  College, University of London, Strand WC2R 2LS, London, U.K.}
\ead{mairi.sakellariadou@kcl.ac.uk} 

\begin{abstract}
We review the noncommutative spectral geometry, a gravitational model
that combines noncommutative geometry with the spectral action
principle, in an attempt to unify General Relativity and the Standard
Model of electroweak and strong interactions. Despite the
phenomenological successes of the model, the discrepancy between the
predicted Higgs mass and the current experimental data indicate that
one may have to go beyond the simple model considered at first. We
review the current status of the phenomenological consequences and
their implications. Since this model lives by construction at high
energy scales, namely at the Grand Unified Theories scale, it provides
a natural framework to investigate early universe cosmology. We
briefly review some of its cosmological consequences.
\end{abstract}

\section{Introduction}
\label{Intro}
Given the plethora of precise cosmological and astrophysical data and
the measurements obtained by particle physics experiments reaching
constantly higher energy scales, we are presently in a position to
falsify early universe cosmological models. In return, by comparing
the theoretical predictions against current data, we are able to
constrain the fundamental theories upon which our cosmological models
were based. A fruitful such example is the cosmological model based on
Noncommutatitive Geometry~\cite{ncg-book1,ncg-book2} and the Spectral
Action principle, leading to the Noncommutative Spectral Geometry
(NCSG), a theoretical framework that can provide~\cite{ccm} a purely
geometric explanation for the Standard Model (SM) of strong and
electroweak interactions. This model lives by construction in high
energy scales, the Grand Unified Theories (GUTs) scale, offering a natural
framework to study early universe
cosmology~\cite{Nelson:2008uy}-\cite{Sakellariadou:2011dk}.  Hence,
instead of postulating a Lagrangian upon which we will build our
cosmological model, we will adopt the one dictated by NCSG, with
the constraints imposed by NCSG itself, and within the framework of
this gravitational theory we will address some cosmological issues.
Clearly, this is a solid approach, since
the cosmological model is inspired and controlled from a fundamental
theory.  In return, by comparing the predictions of the model against
high energy physics measurements and cosmological data, we will be
able to constrain some of the free parameters of NCSG  and/or
its basic element, namely the choice of the algebra.

In this presentation, we will first briefly
review~\cite{Sakellariadou:2012jz} some elements of NCSG and 
then discuss some cosmological consequences of the model. Our purpose
is not to give a full and detailed analysis, but rather  highlight
some results and open questions and  guide the reader through this
powerful interplay between mathematics, high energy physics, cosmology
and astrophysics~\footnote{We refer the reader to the contribution by
Sakellariadou, Stabile and Vitiello~\cite{proc-ssv}, presented in the
same meeting, for a discussion on firstly, the physical meaning of the
choice of the geometry and its relation to quantisation, and secondly,
the relation of NCSG to the gauge structure of the theory and
to dissipation, summarising the results of Ref.~\cite{PRD}.}.

Let us clarify that the approach we will follow here is {\sl a priori}
distinctive from the noncommutative approach based upon $[x^i,
x^j]=i\theta^{ij}$, where $\theta^{ij}$ is an anti-symmetric real
$d\times d$ matrix, used to implement fuzziness of space-time (with
$d$ the space-time dimensionality). One finds in the literature that
noncommutative space is often of Moyal type, involving noncommutative
tori or Moyal planes.  Note however that the Euclidean version of
Moyal noncommutative field theory is compatible with the spectral
triples formulation of noncommutative geometry. This statement holds
for compact noncommutative spaces, while it has been
argued~\cite{gayral} that the compactness restriction is merely a
technical one and noncompact noncommutative spin geometry can be
built, implying that the classical background of noncommutative field
theory can be recast in the spectral triple approach developed by
Connes to describe noncommutative spaces.

\section{Elements of Noncommutative Spectral Geometry}
\label{NCSG}
It is reasonable to argue that the notion of geometry, as we are
familiar with, loses its meaning at very high energy scales, namely
near and above Planck scale. The simple classical picture and the
notion of a continuous space should cease to be valid as quantum
gravity effects turn on. Thus, according to one such school of
thought, one may argue that at sufficiently high energy scales,
spacetime becomes discrete and coordinates no longer commute.  The
noncommutative spacetime can be thus seen as a quantum effect of
gravity, an approach that may shed some light on the regularisation of
quantum field theory. One can, to a first approximation, consider the
simplest class of noncommutative spaces (almost commutative), which
are not incompatible with low energy physics (namely, today's physics)
and study their consistency with experimental and observational data
within the realm of high energy physics and cosmology,
respectively. If such a programme passes successfully this nontrivial
test, as a second step, one should attempt to construct less trivial
noncommutative spaces whose limit is this simple but successful case
studied first. All current studies remain at present within the first
step of this promising programme.

The choice of the noncommutative space, followed by Alain Connes and
his collaborators, was such that at low energy scales one recovers the
SM action. There is an intrinsic difference between this approach and
other ones that attempt to capture the effects of quantum gravity,
which can no longer be switched off once we reach Planck energy
scale. More precisely, in the NCSG approach one does not postulate the
physics at very high energy scales, but instead one is guided by low
energy physics.  Hence, within NCSG the SM is considered as a
phenomenological model which dictates the geometry of space-time so
that the Maxwell-Dirac action functional leads to the SM action.  To
be more specific, in the framework of NCSG we are following here,
gravity and the SM fields are put together into matter and geometry on
a noncommutative space made from the product of a four-dimensional
commutative manifold by a noncommutative internal space. Combining
noncommutative geometry with the spectral action principle, and
choosing the smallest finite dimensional algebra that can account for
the SM particles, Connes and his collaborators have obtained a purely
geometric explanation for the SM Lagrangian coupled to gravity. Thus,
the model we will briefly present here, has been tailored in order to
give the SM of particle physics. This implies that if this model leads
to small discrepancies with respect to the data, one could then first
examine a larger algebra (that could accommodate particles beyond the
SM sector), before considering large deviations from commutative
spaces.

Noncommutative spectral geometry is composed by a two-sheeted space,
made from the product of a smooth four-dimensional manifold ${\cal M}$
(with a fixed spin structure), by a discrete noncommutative space
${\cal F}$ composed by only two points.  The internal space ${\cal F}$
has dimension $6$ to allow fermions to be simultaneously Weyl and
chiral, while it is discrete to avoid the infinite tower of massive
particles that are produced in string theory.  The noncommutative
nature of ${\cal F}$ is given by a spectral triple, introduced by
Connes as an extension of the notion of Riemannian manifold to
noncommutative geometry.  The real spectral triple is given by $\left(
{\cal A}_{\cal F},{\cal H}_{\cal F}, D_{\cal F}\right)$.  The algebra
${\cal A}_{\cal F}=C^\infty({\cal M})$ of smooth functions on ${\cal
M}$ is an involution of operators on the finite-dimensional Hilbert
space ${\cal H_F}$ of Euclidean fermions; it is essentially the
algebra of coordinates.  The operator $D_{\cal F}$ is the Dirac
operator ${\partial\hspace{-5pt}\slash}_{\cal
M}=\sqrt{-1}\gamma^\mu\nabla_\mu^s$ on the spin Riemannian manifold
${\cal M}$. It is such that $J_{\cal F}{\cal D_F} = \epsilon'{\cal
D_F}J_{\cal F}$, where $J_{\cal F}$ is an anti-linear isometry of the
finite dimensional Hilbert space, with the properties $J_{\cal
F}^2=\epsilon~,~J_{\cal F}\gamma_{\cal F}=\epsilon''\gamma_{\cal F}
J_{\cal F}$, with $\gamma_{\cal F}$ the chirality operator and
$\epsilon,\epsilon',\epsilon''\in \{\pm 1\}$.  The operator $D_{\cal
F}$ is a linear self-adjoint unbounded operator and corresponds to the
inverse of the Euclidean propagator of fermions; it is given by the
Yukawa coupling matrix which encodes the masses of the elementary
fermions and the Kobayashi--Maskawa mixing parameters.

The spectral geometry is then given by the product rules:
\be
{\cal A}=C^\infty({\cal M})\oplus{\cal A_F}\ \ ,\ \ 
{\cal H}=L^2({\cal M},S)\oplus{\cal H_F}\ \ , \ \ 
{\cal D}={\cal D_M}\oplus1+\gamma_5\oplus{\cal D_F}~,
\nonumber
\ee
where $L^2({\cal M}, S)$ is the Hilbert space of $L^2$ spinors and
${\cal D_M}$ is the Dirac operator of the Levi-Civita spin connection
on the four-dimensional manifold ${\cal M}$. Note that the chirality
operator is $\gamma=\gamma_5\oplus\gamma_{\cal F}$ and the
anti-unitary operator on the complex Hilbert space is $J=J_{\cal
M}\oplus J_{\cal F}$, with $J_{\cal M}$ the charge conjugation.  

Since in the following we only consider the noncommutative discrete
space ${\cal F}$, we omit the subscript $_{\cal F}$ to keep the
notation lighter.

Assuming the algebra
${\cal A}$ to be symplectic-unitary, it reads~\cite{Chamseddine:2007ia}
\begin{equation}
\mathcal{A}=M_{a}(\mathbb{H})\oplus M_{k}(\mathbb{C})~,
\end{equation}
with $k=2a$; $\mathbb{H}$ is the algebra of quaternions, which encodes
the noncommutativity of the manifold.  The first possible value for
$k$ is 2, corresponding to a Hilbert space of four fermions. This
choice is however ruled out from the existence of quarks. The next
possible value is $k=4$ leading to the correct number of $k^2=16$
fermions in each of the three generations; the number of generations
is a physical input in the theory. The model developed by Connes and
his collaborators is the minimal one ($k=4$) that can account for
the Standard Model.

To obtain the NCSG action we apply the spectral action principle to
the product geometry ${\cal M}\times {\cal F}$.  Thus, the bare bosonic
Euclidean action is simply
\be {\rm Tr}(f(D_{\cal A}/\Lambda))~, \ee
where $D_{\cal A}=D +{\cal A}+\epsilon'J{\cal A}J^{-1}$ are
uni-modular inner fluctuations, $f$ is a cutoff function and $\Lambda$
fixes the energy scale. This action can be seen {\sl \`a la} Wilson as
the bare action at the mass scale $\Lambda$.  The fermionic term can
be included in the action functional by adding $(1/2)\langle
J\psi,D\psi\rangle$, where $J$ is the real structure on the spectral
triple and $\psi$ is a spinor in the Hilbert space ${\cal H}$ of the
quarks and leptons.

Using heat kernel methods, the trace ${\rm Tr}(f({\cal D}_A/\Lambda))$
can be written in terms of the geometrical Seeley-de Witt coefficients
$a_n$, which are known for any second order elliptic differential
operator, as $\sum_{n=0}^\infty F_{4-n}\Lambda^{4-n}a_n$~, where the
function $F$ is defined such that $F({\cal D}_A^2)=f({\cal D}_A)$.  To
be more precise, the bosonic part of the spectral action can be
expanded in powers of $\Lambda$ in the form~\cite{ac1996,ac1997}
\begin{equation}
\label{eq:sp-act}
{\rm Tr}\left(f\left(\frac{{\cal D}_A}{\Lambda}\right)\right)\sim
\sum_{k\in {\rm DimSp}} f_{k} \Lambda^k{\int\!\!\!\!\!\!-} |{\cal
  D}_A|^{-k} + f(0) \zeta_{{\cal D}_A(0)}+ {\cal O}(1)~,
\end{equation}
where $f_k$ are the momenta of the smooth even test (cutoff) function
which decays fast at infinity:
\beq \nonumber 
f_0 &\equiv& f(0)~,\\  \nonumber
f_k &\equiv&\int_0^\infty f(u) u^{k-1}{\rm
  d}u\ \ ,\ \ \mbox{for}\ \ k>0 ~,\nonumber\\ \mbox
    f_{-2k}&=&(-1)^k\frac{k!}{(2k)!} f^{(2k)}(0)~.  \nonumber
\eeq
 The noncommutative integration is defined in terms of residues of
zeta functions $\zeta_{{\cal D}_A} (s) = {\rm Tr}(|{{\cal
D}_A}|^{-s})$ at poles of the zeta function and the sum is over points
in the dimension spectrum of the spectral triple.  For a
four-dimensional Riemannian geometry, the trace ${\rm Tr}(f({\cal
D}_A/\Lambda))$ can be expressed perturbatively
as~\cite{sdw-coeff}-\cite{nonpert}
\be\label{asymp-exp} {\rm Tr}(f({\cal D}_A/\Lambda))\sim
2\Lambda^4f_4a_0+2\Lambda^2f_2a_2+f_0a_4+\cdots
+\Lambda^{-2k}f_{-2k}a_{4+2k}+\cdots~.  \ee
Since the Taylor expansion of the cutoff function vanishes at zero, the
asymptotic expansion of Eq.~(\ref{asymp-exp}) reduces to
\be \label{asympt}
{\rm Tr}(f(D/\Lambda))\sim
2\Lambda^4f_4a_0+2\Lambda^2f_2a_2+f_0a_4~.  \ee
Hence, the cutoff function $f$ plays a r\^ole only through its three
momenta $f_0, f_2, f_4$, which are three real parameters, related to
the coupling constants at unification, the gravitational constant, and
the cosmological constant, respectively. More precisely, the first
term in Eq.~(\ref{asympt}) which is in in $\Lambda^4$ gives a
cosmological term, the second one which is in $\Lambda^2$ gives the
Einstein-Hilbert action functional, and the third one which is
$\Lambda$-independent term yields the Yang-Mills action for the gauge
fields corresponding to the internal degrees of freedom of the metric.

As it has been shown in Ref.~\cite{ccm}, the NCSG summarised above,
offers a purely geometric approach to the SM of particle physics,
where the fermions provide the Hilbert space of a spectral triple for
the algebra and the bosons are obtained through inner fluctuations of
the Dirac operator of the product ${\cal M}\times {\cal F}$ geometry.
More precisely, the computation of the asymptotic expression for the
spectral action functional results to the full Lagrangian for the
Standard Model minimally coupled to gravity, with neutrino mixing and
Majorana mass terms; supersymmetric extensions have been also
considered~\cite{walter}.

\section{Phenomenological consequences}
We will briefly review the phenomenological consequences of the NCSG
model. We assume that the function $f$ is well approximated by the
cutoff function and ignore higher order terms.  Normalisation of the
kinetic terms implies
\be
\frac{g_3^2f_0}{ 2\pi^2}=\frac{1}{4} ~~\mbox{and}~~ g_3^2=g_2^2=
\frac{5}{ 3}g_1^2\nonumber~, \ee
which coincide with the relations obtained in GUTs, while
\be \sin^2\theta_{\rm W}=\frac{3}{8}~,\ee
which is also found  for the SU(5) and SO(10) groups.  Assuming the
validity of the big desert hypothesis, the running of the couplings
$\alpha_i=g_i^2/(4\pi)$ with $i=1,2,3$ is obtained via the
RGE.

Considering only one-loop corrections, hence the $\beta$-functions are
$\beta_{g_i}=(4\pi)^{-2}b_ig_i^3$ with $i=1,2,3$ and
$b=(41/6,-19/6,-7)$, it was shown~\cite{ccm} that the gauge couplings
and the Newton constant do not meet at a point, the error being within
just few percent. The lack of a unification scale implies that the big
desert hypothesis is only approximately valid and new physics are
expected between the unification and today's energy scales. Phrasing
it differently, the lack of a unique unification scale implies that
even though the function $f$ can be approximated by the cutoff
function, there exist small deviations.  On the positive side, the
NCSG model leads to the correct representations of the fermions with
respect to the gauge group of the SM, the gauge bosons appear as inner
fluctuations along the continuous directions while the Higgs doublet
appears as part of the inner fluctuations of the metric.  Spontaneous
Symmetry Breaking mechanism for the electroweak symmetry arises
naturally with the negative mass term without any tuning. The see-saw
mechanism is obtained and the 16 fundamental fermions (the number of
states on the Hilbert space) are recovered. At unification scale
$\Lambda\sim 1.1\times 10^{17}\ {\rm GeV}$, with $g\sim 0.517$, the
Renormalisation Group Equations (RGE) lead to a top quark mass of
$\sim 179 ~{\rm GeV}$.

However, in zeroth order, the model predicts a heavy Higgs mass of
  $\sim 170 ~{\rm GeV}$, which is ruled out by current experimental
  data.  One has though to keep in mind that the Higgs mass is
  sensitive to the value of unification scale, as well as to
  deviations of the spectral function from the cutoff function we have
  considered.  Hence, the Higgs mass should be determined by
  considering higher order corrections and incorporating them to the
  appropriate RGE. One may argue that the reason for which the top
  quark mass is consistent with experimental data while the predicted
  Higgs mass is ruled out, is simply because the top quark mass is
  less sensitive to the ambiguities of the unification scale than the
  Higgs mass is. This may indeed be the case since the bosonic part of
  the action is given by an infinite expansion assuming convergence of
  higher order terms.

In a more drastic modification of the NCSG model at hand, one may
consider a bigger algebra than the one considered so far, which was
chosen so that it leads to the Standard Model particles.  Thus, one
may argue that the discrepancy between the predicted Higgs mass and
the experimental constraints may be resolved by considering models
beyond the SM.  One may construct~\cite{stephan} such a model based on
a minimal spectral triple which contains the SM particles, but it has
also new vector-like fermions and a new U(1) gauge subgroup.  In the
model presented in Ref.~\cite{stephan} it also appears a new complex
scalar field that couples to the right-handed neutrino, the new
fermions and the standard Higgs field. For the case of a nonzero
vacuum expectation value the new scalar and the Higgs field mix and
the mass eigenstates may consist of a light scalar particle with
$m_{H_1}\sim 120$ GeV and a heavy particle with $m_{H_2}\leq 170$ GeV.

More recently, Chamseddine and Connes have argued~\cite{higgs} that
including a real scalar singlet, strongly coupled to the Higgs
doublet, they can accommodate a Higgs mass of order 125 GeV.  This
singlet field is associated with the Majorana mass of the right-handed
neutrino. Even though this singlet is responsible for the breakdown
of the symmetry of the discrete space, it has been neglected in the
original calculations~\cite{ccm} of the phenomenological consequences
of the spectral action. Note that as we have shown in Ref.~\cite{mmm}
this scalar singlet cannot play the r\^ole of the inflaton field and,
at the same time, provide the seeds of temperature anisotropies.

Nevertheless, it is fair to say that even the simplest version of
NCSG predicts a Higgs mass of the correct order of magnitude, which is
certainly a nontrivial result.  Finally, let me note that this
approach to unification does not provide any explanation of the number
of generations, nor leads to constraints on the values of the Yukawa
couplings.

\section{Cosmological consequences}
Since the NCSG gravitational model lives by construction are high
energy scales, it provides a natural framework to investigate early
universe cosmology~\cite{Nelson:2008uy}-\cite{Sakellariadou:2011dk}.
We will review some of these consequences in what follows.

Within NCSG it is natural to obtain the bosonic action in Euclidean
signature. It reads~\cite{ccm}
\beq\label{eq:action1} 
{\cal S}^{\rm E} = \int \left(
\frac{1}{2\kappa_0^2} R + \alpha_0
C_{\mu\nu\rho\sigma}C^{\mu\nu\rho\sigma} + \gamma_0 +\tau_0 R^\star
R^\star
\right.  
+ \frac{1}{4}G^i_{\mu\nu}G^{\mu\nu
  i}+\frac{1}{4}F^\alpha_{\mu\nu}F^{\mu\nu\alpha}\nonumber\\ 
+\frac{1}{4}B^{\mu\nu}B_{\mu\nu}
+\frac{1}{2}|D_\mu{\bf H}|^2-\mu_0^2|{\bf H}|^2
\left.
- \xi_0 R|{\bf H}|^2 +\lambda_0|{\bf H}|^4
\right) \sqrt{g} \ d^4 x~, \eeq
where 
\beq\label{bc} 
\kappa_0^2=\frac{12\pi^2}{96f_2\Lambda^2-f_0\mathfrak{c}}
~~&,&
~~\alpha_0=-\frac{3f_0}{10\pi^2}~~~,\nonumber\\ 
\gamma_0=\frac{1}{\pi^2}\left(48f_4\Lambda^4-f_2\Lambda^2\mathfrak{c}
+\frac{f_0}{4}\mathfrak{d}\right)~~&,&
~~\tau_0=\frac{11f_0}{60\pi^2}~~~,\nonumber\\ 
\mu_0^2=2\Lambda^2\frac{f_2}{f_0}-{\frac{\mathfrak{e}}{\mathfrak{a}}}~~~,
~~~\xi_0=\frac{1}{12}~~~&,&
~~~\lambda_0=\frac{\pi^2\mathfrak{b}}{2f_0\mathfrak{a}^2}~;
\eeq
${\bf H}$ is a rescaling ${\bf H}=(\sqrt{af_0}/\pi)\phi$ of the Higgs
field $\phi$ to normalise the kinetic energy. The geometric parameters
$\mathfrak{a, b, c, d, e}$ correspond to the Yukawa parameters (which
run with the RGE) of the particle physics model and the Majorana terms
for the right-handed neutrinos. Extrapolations to lower energy scales
are possible through RGE analysis, however at low (today's) energy
scales nonperturbative effects can no longer be neglected. Thus, any
results based on the asymptotic expansion and on RGE analysis can only
be valid for early universe cosmology.  Note also that the relations
in Eq.~(\ref{bc}) are tied to the scale of the asymptotic expansion;
there is no reason for these constraints (boundary conditions) to hold
at scales below the unification scale $\Lambda$.

To apply the NCSG action in cosmology we must express it in Lorentzian
signature. Assuming that one can perform a Wick rotation in imaginary
time, the Lorentzian version of the gravitational part of the
asymptotic expression for the bosonic sector of the NCSG action
reads~\cite{ccm}
\be\label{eq:1.5} {\cal S}_{\rm grav}^{\rm L} = \int \left(
\frac{1}{2\kappa_0^2} R + \alpha_0
C_{\mu\nu\rho\sigma}C^{\mu\nu\rho\sigma} + \tau_0 R^\star
R^\star
\xi_0 R|{\bf H}|^2 \right)
\sqrt{-g} \ d^4 x~,\ee
leading to the equations of motion~\cite{Nelson:2008uy}:
\be\label{eq:EoM2} R^{\mu\nu} - \frac{1}{2}g^{\mu\nu} R +
\frac{1}{B^2} \delta_{\rm cc}\left[
  2C^{\mu\lambda\nu\kappa}_{;\lambda ; \kappa} +
  C^{\mu\lambda\nu\kappa}R_{\lambda \kappa}\right]
\nonumber\\ = 
\kappa_0^2 \delta_{\rm cc}T^{\mu\nu}_{\rm matter}~, \ee
where  $B^2 \equiv -(4\kappa_0^2
\alpha_0)^{-1}$.
The nonminimal coupling between the Higgs field
and the Ricci curvature scalar is captured by the parameter $\delta_{\rm cc}$,
defined by $\delta_{\rm cc}\equiv[1-2\kappa_0^2\xi_0{\bf H}^2]^{-1}.$

In the low energy weak curvature regime, the nonminimal coupling
between the background geometry and the Higgs field can be neglected,
leading to $\delta_{\rm cc}=1$.  In this regime, noncommutative
corrections do not occur at the level of a
Friedmann-Lema\^{i}tre-Robertson-Walker (FRLW) background, since in
this case the modified Friedmann equation reduces to its standard
form~\cite{Nelson:2008uy}.  Any modifications to the background
equation will be apparent at leading order only for anisotropic and
inhomogeneous models.  For instance, consider the Bianchi type-V
model, for which the space-time metric, in Cartesian coordinates,
reads
\be g_{\mu\nu} = {\rm diag} \left[ -1,\{a_1(t)\}^2e^{-2nz} ,
 \{a_2(t)\}^2e^{-2nz}, \{a_3(t)\}^2 \right]~; \ee
$a_i(t)$ with $i=1,2,3$ are arbitrary functions, denoting the scale
factors in the three spatial coordinates and $n$ is an integer.
Defining
$A_i\left(t\right) = {\rm ln} a_i\left(t\right)$  with $i=1,2,3$,
the modified Friedmann equation reads~\cite{Nelson:2008uy}:
\beq\label{eq:Friedmann_BV} \kappa_0^2 T_{00}=&&\nonumber\\
 - \dot{A}_3\left(
\dot{A}_1+\dot{A}_2\right) -n^2 e^{-2A_3} \left( \dot{A}_1
\dot{A}_2-3\right)&& \nonumber \\
 +\frac{8\alpha_0\kappa_0^2 n^2}{3} e^{-2A_3} \left[
  5\left(\dot{A}_1\right)^2 + 5\left(\dot{A}_2\right)^2 -
  \left(\dot{A}_3\right)^2\right.
\left. 
- \dot{A}_1\dot{A}_2 - \dot{A}_2\dot{A}_3
  -\dot{A}_3\dot{A}_1 - \ddot{A}_1 - \ddot{A}_2 - \ddot{A}_3 + 3
  \right]
&& \nonumber \\
- \frac{4\alpha_0\kappa_0^2}{3} \sum_i \Biggl\{
\dot{A}_1\dot{A}_2\dot{A}_3 \dot{A}_i
 + \dot{A}_i \dot{A}_{i+1} \left(
\left( \dot{A}_i - \dot{A}_{i+1}\right)^2 -
\dot{A}_i\dot{A}_{i+1}\right)
&& \nonumber \\
 + \left( \ddot{A}_i + \left( \dot{A}_i\right)^2\right)\left[
  -\ddot{A}_i - \left( \dot{A}_i\right)^2 + \frac{1}{2}\left(
  \ddot{A}_{i+1} + \ddot{A}_{i+2} \right)
\right.
\left. 
+ \frac{1}{2}\left(
  \left(\dot{A}_{i+1}\right)^2 + \left( \dot{A}_{i+2}\right)^2 \right)
  \right]
&& \nonumber \\ 
+ \left[ \dddot{A}_i + 3 \dot{A}_i \ddot{A}_i -\left(\ddot{A}_i +
  \left(\dot{A}_i\right)^2 \right)\left( \dot{A}_i - \dot{A}_{i+1} -
  \dot{A}_{i+2} \right)\right]
\left[ 2\dot{A}_i
  -\dot{A}_{i+1}-\dot{A}_{i+2} \right]\Biggr\} \eeq
with $i=1,2,3$; the $t$-dependence of the terms has been omitted
for simplicity.  
Any term containing $\alpha_0$ in Eq.~(\ref{eq:Friedmann_BV}) encodes
a modification from the conventional case.  The correction terms can be divided
into two types.  The first one contains the terms in braces in
Eq.~(\ref{eq:Friedmann_BV}), which are fourth order in time
derivatives and hence, for the slowly varying functions that are usually
considered in cosmology, these corrections can be neglected.  The
second type, which appears in the third line in
Eq.~(\ref{eq:Friedmann_BV}), occurs at the same order as the standard
Einstein-Hilbert terms. However, since this correction term is proportional to
$n^2$, it vanishes for homogeneous versions of Bianchi type-V.  
In conclusion, the corrections to Einstein's equations can
only be important for inhomogeneous and anisotropic
space-times~\cite{Nelson:2008uy}.

The coupling between the Higgs field and the background geometry can
no longer be neglected once we reach energies of the Higgs scale. In
this case, the nonminimal coupling of Higgs field to curvature leads
to corrections to Einstein's equations even for homogeneous and
isotropic cosmological models. To keep the analysis simpler, let us
neglect the conformal term in Eq.~(\ref{eq:EoM2}), so that the
equations of motion are~\cite{Nelson:2008uy}
\be R^{\mu\nu} - \frac{1}{2}g^{\mu\nu}R =
\kappa_0^2\left[\frac{1}{1-\kappa_0^2 |{\bf H}|^2/6}\right]
T^{\mu\nu}_{\rm matter}~. \ee
Thus, $|{\bf H}|$ plays the r\^ole of an effective gravitational
constant~\cite{Nelson:2008uy}.  Alternatively, the nonminimal coupling
of the Higgs field to the curvature can be seen, for static
geometries, as leading to an increase of the Higgs
mass~\cite{Nelson:2008uy}. In the presence of this nonminimal
coupling, the cosmological model built upon the NCSG action share some
similarities with chameleon gravity and dilatonic
cosmology~\cite{Nelson:2008uy}.

The nonminimal coupling between the Higgs field and the Ricci
curvature may turn out to be crucial in early universe
cosmology~\cite{Nelson:2009wr,mmm}.  Such a coupling has been
introduced {\sl ad hoc} in the literature, in an attempt to drive
inflation through the Higgs field. However, the value of this coupling
should be dictated by an underlying theory and cannot be tuned by hand
for a purely phenomenological convenience.  Actually, even if
classically the coupling between the Higgs field and the Ricci
curvature could be set equal to zero, a nonminimal coupling will be
induced once quantum corrections in the classical field theory are
taken into account.  A large coupling between the Higgs field and the
background geometry is plagued by pathologies~\cite{higgs-infl1,
higgs-infl2}, but this is not the case for a small coupling, as in the
NCSG case.  It is therefore worth investigating whether the Higgs
field could play the r\^ole of the inflaton within the NCSG context.

In a FLRW metric, the Gravity-Higgs sector of the asymptotic
expansion of the spectral action, in Lorentzian
signature, reads
\be
S^{\rm
  L}_{\rm GH}=\int\Big[\frac{1-2\kappa_0^2\xi_0
    H^2}{2\kappa_0^2}R 
-\frac{1}{2}(\nabla  H)^2- V(H)\Big] \sqrt{-g}\  d^4x~,
\ee
where 
\be\label{higgs-pot}
V(H)=\lambda_0H^4-\mu_0^2H^2~,
\ee
with $\mu_0$ and $\lambda_0$ subject to radiative corrections as
functions of energy.  For large enough values of the Higgs field, the
renormalised value of $\mu_0$ and $\lambda_0$ must be calculated.  At
high energies the mass term in Eq.~(\ref{higgs-pot}) is sub-dominant
and can be neglected. It has been shown~\cite{mmm} that for each value
of the top quark mass there is a value of the Higgs mass where the
effective potential is about to develop a metastable minimum at large
values of the Higgs field and the Higgs potential is locally
flattened.  Calculating~\cite{mmm} the renormalisation of the Higgs
self-coupling up to two-loops, we have constructed an effective
potential which fits the renormalisation group improved potential
around the flat region.  The analytic fit to the Higgs potential
around the minimum of the potential is~\cite{mmm}:
\be
V^{\rm eff}=\lambda_0^{\rm eff}(H)H^4
=[a\ln^2(b\kappa H)+c] H^4~,
\ee
where the parameters $a, b$ are related to the low energy values of
top quark mass $m_{\rm t}$ as~\cite{mmm}
\beq
a(m_\text{t})&=&4.04704\times10^{-3}-4.41909\times10^{-5}
\left(\frac{m_\text{t}}{\text{GeV}}\right)
+1.24732\times10^{-7}\left(\frac{m_\text{t}}{\text{GeV}}\right)^2~,
\nonumber\\ 
b(m_\text{t})&=&\exp{\left[-0.979261
\left(\frac{m_\text{t}}{\text{GeV}}-172.051\right)\right]}~.
\eeq
The third parameter, $c$, encodes the appearance
of an extremum and depends on the values for top quark mass and Higgs
mass.  
The region where the potential is flat is narrow, thus to achieve a
long enough period of quasi-exponential expansion, the slow-roll parameters,
$\epsilon$ and $\eta$, must be slow enough to allow sufficient number
of e-folds. In addition, the amplitude of density perturbations
$\Delta_\mathcal{R}^2$ in the Cosmic Microwave Background (CMB) must
be within the window allowed from the most recent experimental
data. More precisely, inflation predicts that at horizon crossing
(denoted by stars), the amplitude of density perturbations is related
to the inflaton potential through $\left(V_*/\epsilon_*\right)^{1/4}
=2\sqrt{3\pi}\ m_\text{Pl}\ \Delta_\mathcal{R}^{1/2}$, where
$\epsilon_*\leq1$.  Its value, as measured by
WMAP7~\cite{Larson:2010gs}, requires
$\left(V_*/\epsilon_*\right)^{1/4} =(2.75\pm0.30)\times
10^{-2}\ m_\text{Pl}$, where $m_\text{Pl}$ is the Planck mass.

Performing a systematic search in the parameter space using a
Monte-Carlo chain, we have shown~\cite{mmm} that even though slow-roll
inflation can be realised, the ratio of perturbation amplitudes is too
large for any experimentally allowed values for the masses of the top
quark and the Higgs boson. Note that running of the gravitational
constant and corrections by considering the more appropriate
de\,Sitter space-time, instead of the Minkowski geometry employed
here, do not improve substantially the realisation of a successful
slow-roll inflationary era~\cite{mmm}.

Let us now proceed with the study of linear perturbations around a
Minkowski background metric, which will allow us to constrain one of
the three free parameters of the theory, namely the moment $f_0$ which
is related to the coupling constants at unification. Note that we have
to go beyond an FLRW space-time, since for a homogeneous and isotropic
geometry the Weyl tensor vanishes, implying that the NCSG corrections
to the Einstein equation vanish~\cite{Nelson:2008uy}, rending
difficult to restrict $B$ via cosmology or solar-system tests. It is
worth noting that imposing a lower limit on $B$ would imply an upper
limit to the moment $f_0$, and thus restrict particle physics at
unification.  To impose an upper limit to the moment $f_0$, we will
study the energy lost to gravitational radiation by orbiting
binaries~\cite{Nelson:2010ru,Nelson:2010rt}.  Considering linear
perturbations around a Minkowski background metric, the equations of
motion read~\cite{Nelson:2010rt}
\be\label{eq:1} \left( \Box - B^2 \right) \Box h^{\mu\nu} =
B^2 \frac{16\pi G}{c^4} T^{\mu\nu}_{\rm matter}~, \ee 
where $T^{\mu\nu}_{\rm matter}$ is taken to lowest order in
$h^{\mu\nu}$. Since $B$ plays the r\^ole of a mass, it must be
real and positive, thus $\alpha_0$ must be negative for Minkowski
space to be a stable vacuum of the theory.

Consider the energy lost to gravitational radiation by orbiting
binaries. In the far field limit, $|{\bf r}| \approx |{\bf r} - {\bf
  r}'|$ (${\bf r}$ and ${\bf r}'$ denote the locations of observer
and emitter, respectively), the spatial components of the general
first order solution for a perturbation against a Minkowski background
read~\cite{Nelson:2010rt}
\be\label{eq:4} h^{ik}\left( {\bf r},t\right) \approx \frac{2G
  B}{3c^4} \int_{-\infty}^{t-\frac{1}{c}|{\bf r}|} \frac{d
  t'}{\sqrt{c^2\left( t-t'\right)^2 - |{\bf r}|^2} }
          {\cal J}_1 \left( B\sqrt{c^2\left( t-t'\right)^2 - |{\bf
              r}|^2}\right) \ddot{D}^{ik}\left(t'\right)~; \ee
$D^{ik}$ is the quadrupole moment, defined as $D^{ik}\left(t\right)
\equiv \frac{3}{c^2}\int x^i x^k T^{00}({\bf r},t) \ d{\bf r}$, and
${\cal J}_1$ a Bessel function of the first kind. In the $B\rightarrow
\infty$ limit, the NCSG gravitational theory reduces to the standard
General Relativity (GR), while for finite $B$ gravity waves radiation
contains massive and massless modes, which are both sourced from the
quadrupole moment of the system.

Considering a binary pair of masses $m_1, m_2$ in circular (for
simplicity) orbit in the $(xy)$-plane, the rate of energy loss is
\be\label{eq:energy} -\frac{{\rm d} {\cal E}}{{\rm d}t} \approx
\frac{c^2}{20G} |{\bf r}|^2 \dot{h}_{ij} \dot{h}^{ij}~,  \ee
with~\cite{Nelson:2010rt}
\be
\dot{h}^{ij}\dot{h}_{ij}= \frac{128\mu^2|\rho|^4 \omega^6 G^2
  B^2}{c^8}
 \times \left[ f_{\rm c}^2\left(B|{\bf
    r}|,\frac{2\omega}{B c}\right) + f_{\rm s}^2\left(B|{\bf
    r}|,\frac{2\omega}{B c}\right)\right]~, \ee
\beq\label{eq:f1}
 f_{\rm s}\left( x,z\right) &\equiv& \int_0^\infty
\frac{d s}{\sqrt{s^2 + x^2}} {\cal J}_1\left(s\right) \sin
\left(z\sqrt{ s^2 + x^2} \right)~,\\
\label{eq:f2}
f_{\rm c}\left( x,z\right) &\equiv&
\int_0^\infty \frac{d s}{\sqrt{s^2 + x^2}} {\cal
  J}_1\left(s\right) \cos \left(z\sqrt{ s^2 + x^2} \right)~.
\eeq
The orbital frequency $\omega$, defined in terms of the magnitude
$|\rho|$ of the separation vector between the two bodies, is constant
and equal to $\omega = |\rho|^{-3/2} \sqrt{ G\left( m_1 +
  m_2\right)}$.

The integrals in Eqs.~(\ref{eq:f1}), (\ref{eq:f2}), exhibit a strong
resonance behavior at $z=1$, which corresponds to the critical
frequency~\cite{Nelson:2010rt}
\be
\label{critical}
2\omega_{\rm c} =B c~,
\ee
around which strong deviations from the GR results are expected.  This
maximum frequency results from the natural length scale, given by
$B^{-1}$, at which NCSG effects become dominant.

One can find in the literature several binary pulsars for which the
rate of change of the orbital frequency is well-known and the
predictions of GR agree with the data to a high accuracy. Since the
magnitude of the NCSG deviations from GR must be less than the allowed
uncertainty in the data, we are able to constrain~\cite{Nelson:2010ru}
$B$, namely
\be 
\label{constr-beta}
B > 7.55\times 10^{-13}~{\rm m}^{-1}~.
\ee
This constraint is not too strong but since it is obtained
from systems with high orbital frequencies, future observations
of rapidly orbiting binaries, relatively close to the Earth, could
improve it by many orders of magnitude.

\section{Conclusions}
We have briefly described the prescription of Connes and collaborators
in order to recover the action of the SM of particle physics from
purely gravitational considerations. In this approach one recovers the
Einstein plus Yang-Mills and Weyl actions including the spin-1 bosons
and the part induced by the spin-0 Higgs fields. Thus, gravity and the
electroweak and strong forces are described as purely gravitational
forces on a unified noncommutative spacetime. The trick consists of
employing spectral triples, consisting of an algebra (that is
equivalent of a topological space), a Dirac operator (that corresponds
to the metric on the topological space) and a Hilbert space of Dirac
4-spinors, on which the algebra is represented and on which the Dirac
operator acts. In his approach, Connes combined spacetime with an
internal space, composed by only 2 points, a construction that can be
seen as a discrete Kaluza-Klein space where the product manifold of
spacetime with extra spatial dimensions is replaced by the product of
spacetime with discrete spaces represented by matrices.

The phenomenological consequences of this gravitational theory are in
a very good agreement with current data, however the predicted Higgs
mass seems to be ruled out. The reason for this discrepancy may be due
to the ambiguities of the unification scale, the approximation of the
spectral function by a cutoff one, or the assumption of convergence of
the higher order terms in the infinite expansion of the bosonic part
of the NCSG action. A more drastic approach is to consider larger
algebras which predict particles beyond the SM sector. Even though
these directions deserve further investigations, it is clear that the
NCSG approach offers a beautiful explanation of the SM, the most
successful particle physics model we have at hand, providing in
addition a geometric explanation for the Higgs field which is
otherwise introduced by hand.

The NCSG action offers a natural framework to build a cosmological
model, which may allow us to explain some open questions in cosmology,
as for instance through investigations on the r\^ole of scalar fields
which arise naturally in NCGS and they do not have to be introduced by
hand. Moreover, comparison of NCSG predictions against astrophysical
data may allow us to constrain the free parameters of the theory.

Even though, as we have shown, the Higgs field can play the r\^ole of
the inflaton field, in terms of providing a slow-roll period of fast
expansion, it cannot provide the seed of perturbations leading to the
observed CMB anisotropies and the large-scale structure. If one wants
to decouple the inflaton from the curvaton field, then one has to
provide such a field with the appropriate potential. In this sense,
one may investigate the r\^ole of other scalar fields within the NCSG
model.  It is particularly encouraging that by studying gravity waves
propagation we were able to constrain one of the free parameters of
the theory, namely the one related to the coupling constants at
unification.

In conclusion, NCSG offers a fruitful interplay between mathematics,
gravitational theories, particle physics and cosmology, tracing
another approach to the goal of unification.

\end{document}